\documentclass[twocolumn,aps,floatfix,psfig,superscriptaddress,epsf]{revtex4-1} 
\usepackage{amssymb,amsmath} 
\usepackage{color}
\usepackage{multirow}
\usepackage{braket}
\usepackage{tabularx,booktabs}  
\usepackage{graphicx}
\usepackage{dcolumn}
\usepackage{placeins}
\usepackage{bm}
\usepackage{upgreek}
\usepackage{marvosym }
\usepackage{mathtools}
\usepackage{ulem} 
\usepackage{float}
\usepackage{latexsym,epsfig}  
\newcommand{\be}{\begin{equation}}
\newcommand{\ee}{\end{equation}} 
\newcommand{\bea}{\begin{eqnarray}}   
\newcommand{\eea}{\end{eqnarray}}  
 
\usepackage{comment}

\begin{document}
\title{
Chiral orbital/spin textures and Edelstein effects in monolayer Janus TMDs\\
}
\author{Pratik Sahu}
\email{pratiksahu2413@gmail.com}
\affiliation{Center for Atomistic Modelling and Materials Design, Indian Institute of Technology Madras, Chennai 600036, India } 
\affiliation{Condensed Matter Theory and Computational Lab, Department of Physics, Indian Institute of Technology Madras, Chennai 600036, India }

\author{S. Satpathy}
\email{satpathys@missouri.edu}
\affiliation{Department of Physics \& Astronomy, University of Missouri, Columbia, MO 65211, USA} 
\affiliation{Condensed Matter Theory and Computational Lab, Department of Physics, Indian Institute of Technology Madras, Chennai 600036, India }

\author{B. R. K. Nanda}
\email{nandab@iitm.ac.in}
\affiliation{Center for Atomistic Modelling and Materials Design, Indian Institute of Technology Madras, Chennai 600036, India }
\affiliation{Condensed Matter Theory and Computational Lab, Department of Physics, Indian Institute of Technology Madras, Chennai 600036, India }

\begin{abstract}
We investigate the orbital and spin Edelstein effect(OEE and SEE) in two-dimensional Janus transition metal dichalcogenides (TMDs) of the form MXX$^\prime$ $(M = Mo,\ W,\ Nb;\ X/X^\prime = S,\ Se,\ Te)$ with the aid of density functional theory calculations and tight-binding model Hamiltonian studies. The chalcogen layers $X$ and $X^\prime$, break the mirror symmetry to introduce an internal electric field $E_{int}$ normal to the plane, which is responsible for OEE and SEE. Our results show that in a  non-Janus framework, the wavefunctions at the valence and conduction bands are dominated with the $\ket{x^2-y^2}$, $\ket{xy}$, and $\ket{z^2}$ orbitals. Due to the $E_{int}$ of the Janus system, these orbitals are now intermixed with the $\ket{xz}$ and $\ket{yz}$ orbitals to produce a robust orbital texture around the valleys $\Gamma,K$ and $K^\prime$. The spin orbit coupling, in addition to the formation of a spin texture, introduces a chirality reversal to the orbital texture. An applied in plane electric field creates both OEE and SEE with the former being one order higher in magnitude. This makes the Janus materials promising for spin-orbitronics. Our work paves the way for further experimental exploration for orbital and spin orbital torque in Janus TMDs.

\end{abstract}   
\date{\today}			 		
\maketitle
\section{Introduction}
\textcolor{black}{Manipulation of the orbital degree of freedom for information processing, commonly known as orbitronics, has been drawing increasing attention in the recent years\cite{ohe-1,Go,Jo,pratik-ohe,sayantika_1,sayantika_2,ot-1,ot-2,ot-3,rap1,rap2}.
Orbitronics is an emerging topic and possesses tremendous potentials for tunable quantum transport and devices, and with right symmetries, both magnetic and non-magnetic systems can be potential host for orbitronics. Concurrently, an wider research domain of spin-orbitronics is evolving, where both spin and orbital angular momentum are exploited for the quantum transport. 
Interestingly, it has been found that many of the spin driven phenomena such as spin Hall effect\cite{she-2,she-3}, spin orbit torque\cite{ot-1,ot-2,ot-3,pranaba}, spin Rashba effect\cite{rashba}, etc., partially or fully originate from orbital angular momentum through spin orbit coupling (SOC)\cite{Jo,pratik-ohe,sayantika_1}.}

The orbital transport is fundamentally governed by the orbital texture,  which is defined as the arrangement of orbital angular momenta in the momentum space\cite{Go}. The crystal symmetries have a deterministic effect on the orbital texture. As an example, it has been shown by many including us that broken inversion\cite{pratik-ohe,sayantika_1,sayantika_2,rap1,Go} and time-reversal symmetry\cite{pratik-ohe-mag,pranaba} can significantly enhance the orbital and spin Hall effect.  

One of the interesting quantum phenomena demonstrated both by spin and orbital textures is the Rashba-Edelstein effect\cite{rashba,edel-1,oee-1,oee-2}, where the Rashba field generated spin/orbital textures respond to an applied in plane electric field by shifting the Fermi surface and in that process \begin{figure}[H]
    \centering
    \includegraphics[width=0.5\textwidth]{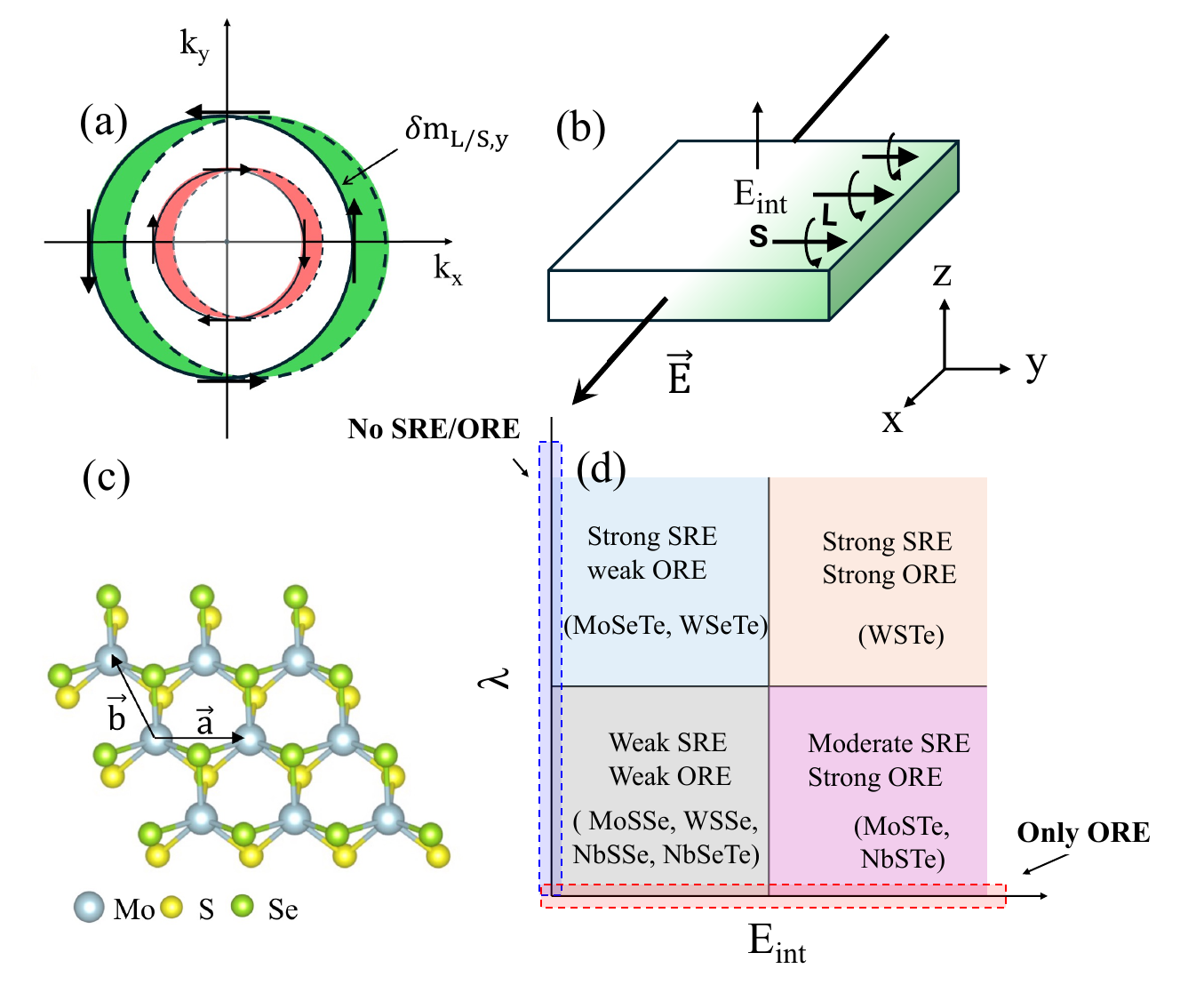}
    \caption{(a) A schematic of the Edelstien effect driven by orbital and spin textures in the momentum space due to an external electric field applied along $\hat{x}$. The green and red regions indicate the accumulation of positive and negative magnetic moments respectively. (b) The magnetoelectric response in the sample due to orbital Edelstein effect (OEE) and spin Edelstein effect(SEE), where $E_{int}$ is the internal electric field, $\vec{E}$ is the applied electric field, and the spin (arrows) and orbital (curved arrows) moments are accumulated on one side of the material. (c) Structure of the monolayer Janus MoSSe compound. The S and Se atoms are in the bottom and top layers respectively, which breaks the mirror symmetry and in turn creates an internal electric field. The lattice translation vectors are shown by $\vec{a}$ and $\vec{b}$. (d) An illustration with several domains for the Janus TMDs, comparing the strength of ORE and SRE, following the strength of the Rashba coefficients listed in Table II. A larger Rashba coefficient would typically lead to a larger Edelstein effect.
    }
    \label{fig:schematic}
\end{figure}\noindent magnetization is developed as shown in Fig. \ref{fig:schematic}(a) and (b). The spin Rashba effect (SRE), usually referred to simply as the Rashba effect,  is characterized by a linear split of the bands and a chiral spin texture in the momentum space\cite{rashba}. Later, it was demonstrated by Park et al.,\cite{park} that asymmetric charge distribution responsible for the Rashba field, gives rise to the orbital Rashba effect (ORE), where a chiral orbital texture forms in the momentum space similar to the spin texture in the case of SRE. Thus, in contrast to the SRE, the ORE does not require the presence of a SOC.
Furthermore, as we realize that orbital transport is as important as spin transport for future devices, it is natural to carry out an in-depth study on the cause and tunability of ORE and identify the host materials possessing it.

There are extensive theoretical and experimental studies showing the ORE on surfaces\cite{ore-1,ore-2,ore-3}, since the latter produce an electric field due to broken mirror symmetry. However, it is not necessary for them to have a strong SOC. On the other hand, the widely investigated 2D-transition metal dichalcogenides(TMDs), $MX_2,$where $M$ is the transition metal atom and $X$ is the chalcogen atom, often possess large SOC and have surface like behavior due to van der Waal separation among the layers. As most of them possess mirror symmetry, the out of plane intrinsic electric field is absent in these materials. In a recent study\cite{tapesh}, a theoretical model was developed to show that a vertical electric field, when applied on pristine TMDs, breaks the mirror symmetry to produce ORE and SRE.
The class of Janus TMDs ($MXX^\prime$), Fig. \ref{fig:schematic}(c),  can naturally create such a vertical electric field as the layer of $X$ and $X^\prime$ have different charge distribution. Despite being promising, although the SRE has been well explored in these Janus systems\cite{sre-janus-1,sre-janus-2,sre-janus-3,rashba-const}, there has been no studies examining the ORE and OEE.

In this work, we investigate the momentum space orbital and spin textures in the Janus TMDs (MXX$^\prime$; M = Mo, W, or Nb and X/X$^\prime$ = S, Se or Te) using density functional theory calculations and Wannier based tight-binding (TB) model. Our studies reveal that these materials show both spin and orbital Edelstein effects. Interestingly, the orbital Edelstein effect, is present in the system with or without the SOC. However, there is a chirality reversal of the orbital texture, when SOC is introduced to the Hamiltonian. 

We find that based on the strength of the internal electric field and SOC, Janus materials can be segregated into four categories, as shown in the schematic phase diagram of Fig. \ref{fig:schematic}(d). It is important to highlight that 
 a finite ORE can be predicted even in the absence of SOC, as long as there is an internal electric field present. Based on the results obtained in this study, WSTe shows strong ORE and SRE, whereas in MoSTe and NbSTe, the ORE is strong and SRE is moderate. A weak ORE with reasonably large SRE is found 
 in MoSeTe and WSeTe. Both SRE and ORE are weak for MoSSe, WSSe, NbSSe, and NbSeTe.

\section{Electronic structure of the Janus materials and formulation of the tight-binding model}
\begin{figure}[!b]
    \centering
\includegraphics[width=0.5\textwidth]{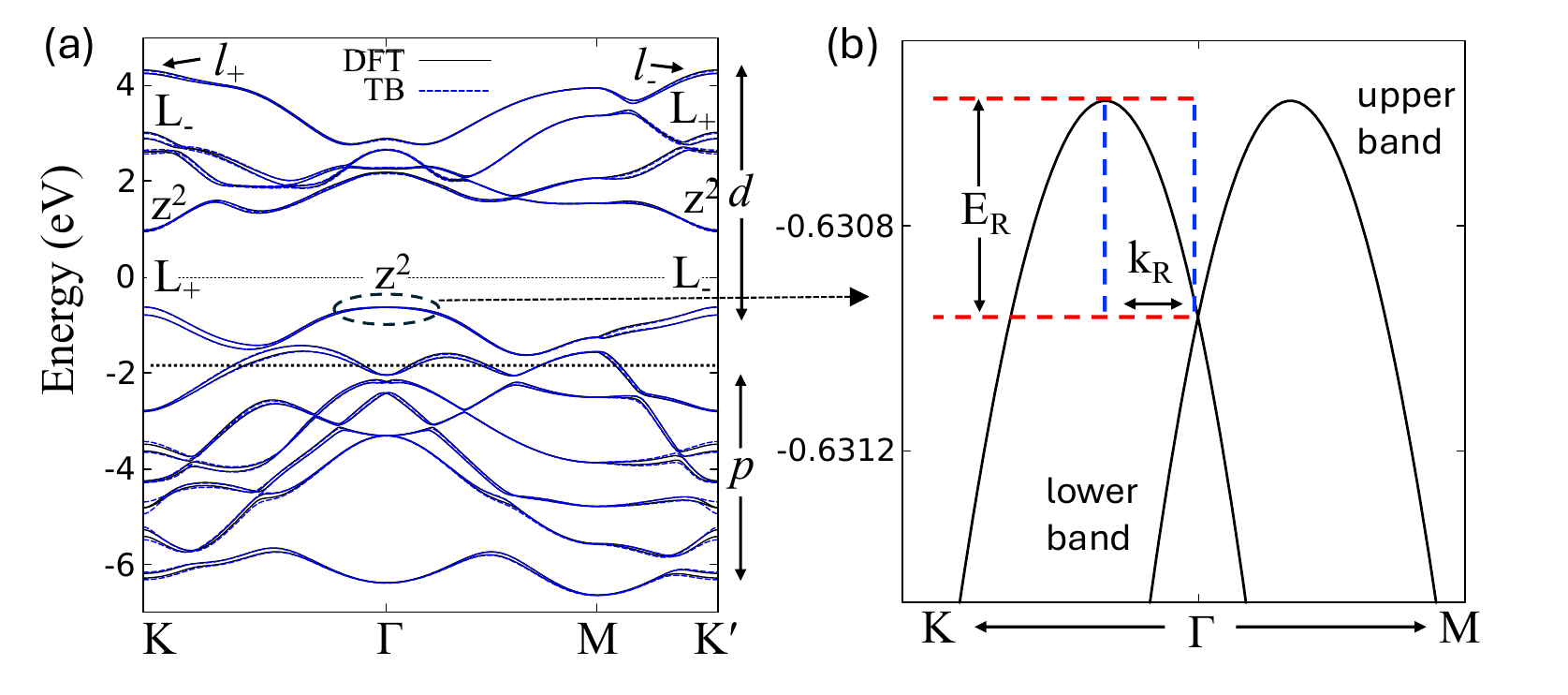}
    \caption{Band structure of monolayer MoSSe in the presence of SOC showing a good match between DFT and the TB model. Here, we have defined the mid gap energy as the zero of the energy. The dominant orbital characters at high-symmetry points for different bands are indicated. Here, $L_\pm$ represents the orbitals $d_{x^2-y^2}\pm i d_{xy}$ and $l_{\pm}$ denotes the orbitals $d_{xz}\pm i d_{yz}$. Due to hybridization and internal mixing, there is a smooth variation in the orbital characters as a function of k-point for each of the bands. The Rashba-split is observed at the high-symmetry points of the bands dominated with $z^2$ character.(b) Zoomed in valence band edge near $\Gamma$, demonstrating the prominent Rashba-split, which is quantified by Rashba-energy split $E_R$, and momentum split $k_R$ respectively. Similar Rashba-splits are observed for the  other Janus materials, and the band structures are provided in the supplemental material\cite{suppl}.
    }
    \label{fig:bands}
\end{figure}

The electronic structure of $MXX^\prime$ is calculated using DFT and further analyzed using Wannier\cite{Wannier90} based tight-binding (TB) model. For the DFT calculations, we have used the norm-conserving Perdew, Burke, and Ernzerhof (PBE) exchange-correlation functionals\cite{PBE} as implemented in Quantum Espresso code\cite{QE}. The self consistency calculation was carried out in a 30$\times$30$\times$1 k-mesh. The DFT+SOC obtained orbital projected band structure of MoSSe, as an example, is shown in Fig. \ref{fig:bands}(a). The top valence bands and the bottom conduction bands are dominated by the Mo-d characters. However, the significance presence of the S-p and Se-p characters suggests that these bands are driven by the d-p hybridizations. The band structures for the rest of the Janus materials are provided in the supplemental material\cite{suppl}.

\textcolor{black}{The band structure of the Janus TMDs is quite similar to the standard TMDs, except for the Rashba splitting of the $z^2$ dominated bands at high-symmetry points. The Rashba-split of the edge valence bands near $\Gamma$, which is one of the important point of this study, is amplified and shown in Fig. \ref{fig:bands}(b). Around this point in the BZ, the internal electric induces the mixing of the $z^2$ orbital with the $xz$ and $yz$ orbitals, which creates the orbital moments along x and y. Since, at the valley points $K/K^\prime$, the finite $\braket{L_z}$ induces a Zeeman type splits of the bands in presence of SOC, the Rashba-splitting is not observed at these k-points. However, a chiral orbital and spin texture can still be observed around these valleys, indicative of the Rashba effect. The orbital and spin moments are computed and discussed in the later sections. This observation is extended to all the compounds mentioned in this work.}

Each of the occupied bands contributes to the orbital texture. Therefore, the TB model is constructed with 11 orbitals in the basis consisting of the Mo-d  and X/X$^\prime$-p orbitals. The hopping parameters defining the bonding among these orbitals were obtained from the Wannier90 program\cite{Wannier90}. The TB Hamiltonian is given by
\begin{equation}
       {\cal {H}} 
 = \sum_{i m\sigma} \varepsilon_{m}c^{\dagger}_{im\sigma}c_{im\sigma}+\sum_{ijmn\sigma}t_{im,jn}c^{\dagger}_{im\sigma}c_{jn\sigma}\\
       +  \frac{2\lambda}{\hbar^2}\vec{L}\cdot\vec{S} ,
    \label{ham}
\end{equation}
where $c^\dagger_{i m\sigma}$ is the electron creation operator with $i/ m/ \sigma$ being the site/orbital/spin index. The first term is the on-site energy, the second term represents the kinetic energy(${\cal {H}}_{KE}$) and, they can be expressed as the following, in the Bloch function basis.
\begin{equation}
    {\cal {H}}_{KE}^{mn} (\vec k) = \sum_{j}t_{mn}^j e^{i\vec{k}\cdot\vec{d_j}},
\end{equation}
where, $t_{mn}^j = \braket{0,m|{\cal {H}}|d_j,n}$ denotes the hopping energy between the $m^{th}$ orbital in the central cell and the  $n^{th}$ orbital located at distance $\vec{d_j}$ with respect to the position of the $m^{th}$ orbital, and the third term corresponds to the SOC(${\cal {H}}_{SO}$). The electron spin is  $\vec{S} = (\hbar/2)\vec{\sigma}$, where $\vec{\sigma}$ are the Pauli matrices.
The orbital angular momentum operators for the case of $L = 1$, in the basis set $\phi_p \equiv (p_x,\ p_y,\ p_z)$ are
\begin{eqnarray}
    &L_x^{(p)} = \hbar\begin{bmatrix}
        0&0&0\\
        0&0&-i\\
        0&i&0
    \end{bmatrix},
    &L_y^{(p)} = \hbar\begin{bmatrix}
        0&0&i\\
        0&0&0\\
        -i&0&0
    \end{bmatrix},
\\ &L_z^{(p)} =\hbar \begin{bmatrix} \nonumber
        0&i&0\\
        i&0&0\\
        0&0&0
    \end{bmatrix},
    \label{l_p}
\end{eqnarray}
while for  $L=2$, with the d-orbital basis set $\phi_d \equiv (z^2, x^2-y^2, xy, yz, xz)$, they are 
\begin{equation}
\begin{split}
    &L_x^{(d)} =\hbar \begin{bmatrix}
        0&0&0&\sqrt{3}i&0\\
        0&0&0&i&0\\
        0&0&0&0&-i\\
        -\sqrt{3}i&-i&0&0&0\\
        0&0&i&0&0
    \end{bmatrix}, \nonumber\\
    &L_y^{(d)} = \hbar\begin{bmatrix}
        0&0&0&0&-\sqrt{3}i\\
        0&0&0&0&i\\
        0&0&0&i&0\\
        0&0&-i&0&0\\
        \sqrt{3}i &-i &0 &0&0
    \end{bmatrix},\\
    &L_z^{(d)} = \hbar\begin{bmatrix}
        0&0&0&0&0\\
        0 &0 &-2i &0 &0\\
        0 &2i &0 &0 &0\\
        0 &0 &0 &0 &i\\
        0 &0 &0 &-i &0
    \end{bmatrix}.\nonumber
    \label{l_d}
\end{split}
\end{equation}
  

The atomistic SOC strengths, $\lambda$, for the compounds were extracted by matching the TB bands with the DFT bands in the presence of SOC and are listed in Table 1. From the table, we find 
that both the transition metal(M) and the chalcogen(X) atoms contribute to the SOC. In addition, we also notice that the p-orbitals of the chalcogens have a stronger SOC than the d-orbitals of several M atoms, which is in agreement with the previous reports\cite{soc-parameter,zoran} based on single atoms.

\begin{table}[t]
    \centering
    \begin{tabular}{c|c|c|c|c|c|c}
    \hline\hline
    Elements & Mo (d) & W (d) & Nb (d) & S (p) & Se (p) & Te (p)\\
    \hline
    $\lambda$ (eV) &0.074 & 0.2 & 0.064& 0.18 & 0.31&0.52\\
    \hline
    atomic value& 0.091 & 0.31 & 0.069 & 0.079 & 0.34 & 0.71\\
    \hline\hline
    \end{tabular}
    \caption{The values of spin orbit coupling constants for different elements obtained by fitting the TB band structure with DFT results. The second row shows the value of $\lambda$ reported in the literature\cite{soc-parameter} for single atoms.}
    \label{tab:my_label}
\end{table}

\section{Orbital and spin texture}
In this section, we examine the orbital and spin-turnings around the $\Gamma,K,$ and $K^\prime$ valleys, and the role of orbital hybridization and SOC, in controlling their behavior. 
The orbital and spin magnetic moments for the occupied bands are given by the semiclassical expression,
\begin{equation}
\begin{aligned}
\vec{m}(\vec{k}) &=\vec{m}_L(\vec{k})+\vec{m}_S(\vec{k})\\
&=\sum _{n}^{occ} \big[\frac{e}{2\hbar} \operatorname{Im} 
\left\langle 
\nabla_{\vec{k}} u_{n\vec k} \middle| 
\big[\mathcal{H}(\vec{k}) - \varepsilon_{n\vec{k}}\big] 
\middle| \nabla_{\vec{k}} u_{n\vec k}
\right\rangle
\\
&+\frac{e}{\hbar} \operatorname{Im}
\left\langle 
\nabla_{\vec{k}} u_{n\vec k} \middle| 
\big[\varepsilon_{n\vec{k}} - E_F\big] 
\middle| \nabla_{\vec{k}} u_{n\vec k})
\right\rangle \\
&-\frac{e}{m_e}\braket{u_{n\vec k}|\vec{S}|u_{n\vec k }}\big] ,
\end{aligned}
\label{eq:L}
\end{equation}
where, $\varepsilon_{n\vec{k}}$, and $u_{n\vec{k}}$ are the eigenvalue and eigenstate for the $n^{th}$ band respectively, and $-e<0$ is the electronic charge.
Here, the first term corresponds to the intrinsic orbital moment, whereas the second term is due to the field dependence of the density of states\cite{Niu}. The second term is an equilibrium property and does not contribute in non-equilibrium situations such as the orbital transport. In the context of the Edelstein effect, which is a Fermi surface property, the second term is zero. In this work, we compute the intrinsic orbital moments by simply using the atom centered approximation (ACA), where the expectation value of the $\vec{L}$ operator is computed. The ACA is often used in the literature\cite{Go, ohe-1, aca-1}, which is thought\cite{intra_inter} to be an excellent approximation in solids with large angular-momentum orbitals, such as the $d$ orbitals here.

First, we focus on the pure orbital effect, and consider the case where SOC is absent. Later, we introduce the SOC and study the behavior of both ORE and SRE. 


\subsection{Orbital Rashba effect: Without SOC}
 In the absence of SOC, the Rashba-spin split does not occur, and we have pair degenerate bands in these non-magnetic systems. In such a case, the spin moments vanish to form any spin texture. However, the polarized field due to the broken mirror symmetry still exists and creates orbital mixing, which generates an orbital texture. Phenomenologically, the formation of orbital texture can be explained as follows. If $\vec{E}_{int}$ is the polarized field induced by the broken mirror symmetry, then an electron moving along the x-direction experiences an effective magnetic field, $\vec{B}_{eff} = -(\hbar/{mc^2} )\ \vec{q}\times\vec{E}_{int}$ along the $\hat{y}$ direction, and the angular momentum $\vec{L}$ of this electron aligns in parallel with $\vec{B}_{eff}$ to produce the orbital texture in the k-space. 
 
\begin{figure}[!htb]
    \centering
    \includegraphics[width=\linewidth]{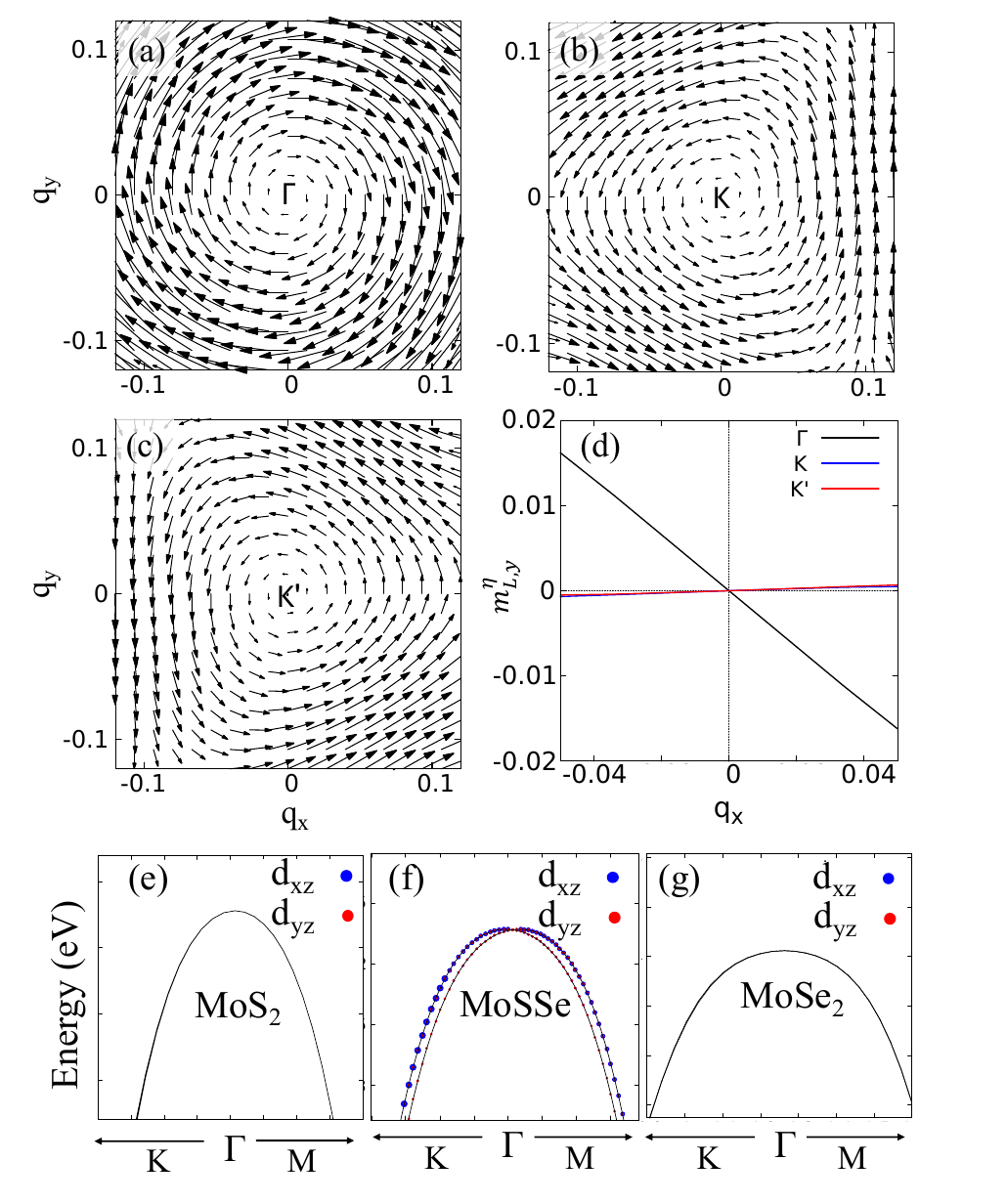}
    \caption{Orbital texture of the MoSSe near (a) $\Gamma$, (b) K and (c) $K'$ valleys, without the SOC. Here, the length of the vectors are scaled in each of the cases for proper visualization of the orbital texture. Importantly, the $\braket{\vec{L}}$ is much stronger near $\Gamma$ as compared to $K/K^\prime$.  (d) The y-component of the orbital moment vector ($m_{L,y}^\eta$) is computed as a function of the small momentum $q_x$ around these valleys $\eta$. The results for the rest of the materials are shown in the supplementary materials\cite{suppl}. The distribution of $d_{xz}$ and $d_{yz}$ orbital characters near $\Gamma$, for (e) MoS$_2$, (f) MoSSe, and (g) MoSTe, highlighting the orbital mixing.
    }
    \label{fig:orb_tex_nosoc}
\end{figure}

 The orbital textures for MoSSe near the valley points are shown in Fig. \ref{fig:orb_tex_nosoc}.  
The orbital textures around the valley points $K$ and $K^\prime$ have chiralities opposite to those near $\Gamma$. Additionally, the orbital moments near $\Gamma$ are found to be much stronger than the rest of the two, suggesting that the $\Gamma$ valley can have a dominant effect on the magnetoelectric response. We also observe that near all valleys the length of the orbital moment vectors varies linearly with the momentum $\vec{q}$ measured with respect to the valley-center, which is a direct result of the linear k-dependence of $\vec{B}_{eff}$. This serves as an important signature of the ORE, and signifies its strength.
Additionally, the orbital moments exhibit a chiral texture, similar to well-known spin Rashba spin texture\cite{rashba-const}. Considering all the features mentioned above, we characterize the orbital texture, by expressing the orbital moments in a simple form, which is

\begin{eqnarray}
    &m_{L,x}^\eta = -\alpha_L^{\eta} \, q \sin\theta=-\alpha_L^\eta q_y, \quad \\
    &m_{L,y}^\eta =\ \alpha_L^{\eta} \, q \cos\theta = \alpha_L^\eta q_x,
    \label{eq_orb_mom}
\end{eqnarray}
where $m^\eta_{L,x}$ and $m^\eta_{L,y}$ are the x and y components of the orbital moments corresponding to the valley $\eta$ ($\Gamma/K/K^\prime$), ($q_x,q_y$) denotes the momentum measured from the valley-center, $\theta$ is the angle made by $\vec{m}_L$, with the x-axis, and $\alpha_L^\eta$ is the slope representing the linear k-dependence of the orbital moments for the valley $\eta$. 

\begin{table}[!htb]
\centering
\setlength{\tabcolsep}{4pt}
\renewcommand{\arraystretch}{1.2}
\begin{tabular}{l|ccc|l|ccc}
\hline\hline
 & $\alpha_L^{\Gamma}$ & $\alpha_L^{K}$ & $\alpha_L^{K'}$ &
 & $\alpha_L^{\Gamma}$ & $\alpha_L^{K}$ & $\alpha_L^{K'}$ \\
\hline
MoSSe  & -0.33 & 0.013 & 0.013 & 
WSSe   & -0.32 & 0.012 & 0.013 \\
MoSTe  & -2.20 & 0.032  & 0.032  & 
WSTe   & -2.34 & 0.063 & 0.063 \\
MoSeTe & -0.14    & 0.005     & 0.005     & 
WSeTe  & -0.12     & 0.005     & 0.005     \\
\hline\hline
\end{tabular}

\vspace{4pt}

\begin{tabular}{l|ccc}
\hline\hline
 & $\alpha_L^{\Gamma}$ & $\alpha_L^{K}$ & $\alpha_L^{K'}$ \\
\hline
NbSSe  & -0.31 & 0.016 & 0.016 \\
NbSTe  & -1.04 & 0.013 & 0.013 \\
NbSeTe & -0.12    & 0.003     & 0.003     \\
\hline\hline
\end{tabular}
\caption{
     The magnitudes of orbital Rashba constant extracted using Eq. \ref{eq_orb_mom} from the variation of $\braket{\vec{L}}$ as a function of $\vec{q}$ for the valence bands. The units are in ($\mu_B\!\cdot\!\text{\AA}$). A specific case is shown in Fig. \ref{fig:orb_tex_nosoc}(d) for MoSSe near $\Gamma$.}
    \label{tab:orb_rashba}

\end{table}
\begin{figure*}[!htb]  
\centering
\includegraphics[width=0.95\textwidth]{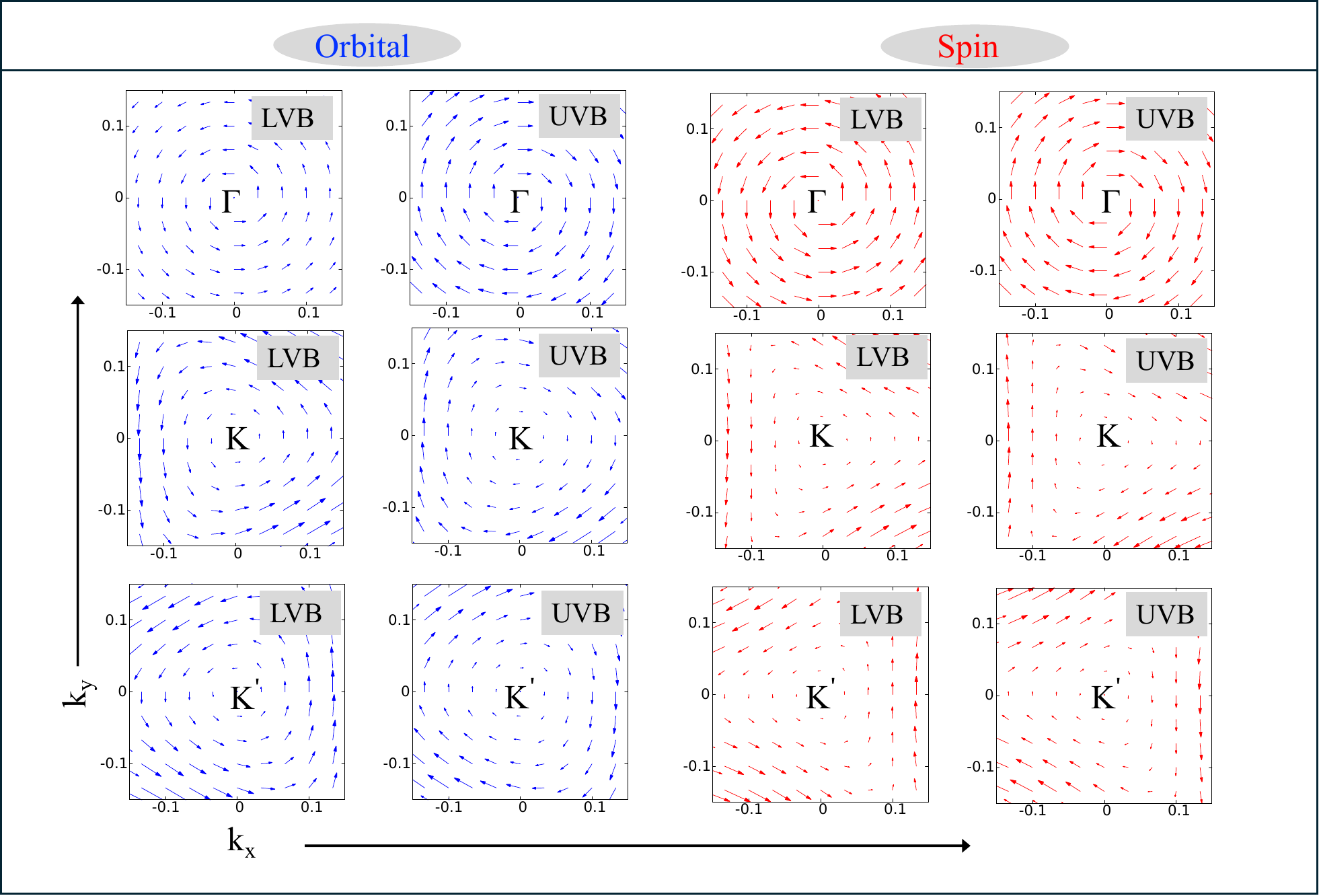}
\caption{Orbital (blue) and spin textures (red) around $\Gamma,K$, and $K^\prime$ valleys in the presence of SOC for monolayer MoSSe. The magnetic moments are computed for the lower and upper Rashba splitted bands separately, which are represented as "L" and "U". The length of the vectors are scaled for visual clarity, with a scaling factor of 50 for the $K$ and $K^\prime$ valleys as compared to the ones for $\Gamma$, which clearly suggests that both the orbital and spin Rashba effects are more prominent near the $\Gamma$ valley. 
}
    \label{fig:moments}
\end{figure*}

Unlike SRE, where the strength is measured in terms of spin Rashba constant $\alpha_R $ $(=\frac{2E_R}{K_R}$, see Fig. \ref{fig:bands}(b)), in the case of ORE, its strength is measured in terms of the slope $\alpha_L^\eta$, and we define this quantity as the orbital Rashba constant. For the case of MoSSe, linear dependence between $m_{L,y}^\eta$ and $\vec{q}$ for $\theta=0$ is shown in Fig.\ref{fig:orb_tex_nosoc}(d).
From Fig.\ref{fig:orb_tex_nosoc}, we gather that the orbital Rashba strength is significantly larger near the $\Gamma$ valley, as compared to the $K/K^\prime$. This is due the fact that, the valence bands near the $\Gamma$ valley is dominated by $d_{z^2}$ orbitals, and Eq. \ref{l_d} demands that in order to have a nonzero $\braket{L_x}$ or $\braket{L_y}$, the eigenstates must have significant contributions from $d_{xz}$ or $d_{yz}$.  In contrast, it is negligible in the valence bands of $K/K'$ valleys. Furthermore, such an orbital mixing is unique to the Janus compounds, as $E_{int}$ is not present in standard TMDs, and hence no such mixing is observed. A comparison of the $d_{xz}$ and $d_{yz}$ characters among three compounds $MoS_2,MoSe_2$ and $MoSSe$ is shown in Fig. \ref{fig:orb_tex_nosoc}(e)-(g).
Comparing the orbital Rashba constants among the Janus compounds, listed in Table. \ref{tab:orb_rashba}, we find that the MSTe compounds show much stronger ORE than the MSSe compounds (see Table \ref{tab:orb_rashba}). This occurs as a result of a larger dipole moment that determines the electric field $E_{int}$. 
\subsection{Orbital and spin Rashba effect: With SOC}
In this section, we examine the orbital and spin moments as we include SOC in our TB Hamiltonian in Eq \ref{ham}. With SOC, many of the bands show a Rashba split, where two degenerate bands split in the momentum space, as discussed earlier in the context of Fig. \ref{fig:bands}. 
The orbital textures near the valleys for the Rashba-split resultant upper and lower bands are plotted in Fig. \ref{fig:moments}. The orbital texture formed by the upper band has a clockwise orientation, while the lower band has an anti-clockwise orientation around $\Gamma$. This is in contrast with the pair of non-SOC degenerate bands, whose orbital textures have clockwise orientation as we discussed in the previous section. As a whole, with SOC, we see a chirality reversal of the orbital textures in the neighborhood of each valley.

%

Along with the chirality reversal of the orbital texture, the SOC now introduces a non-zero chiral spin texture as shown in Fig. \ref{fig:moments}. As typical for the spin Rashba effect, the spin turnings for the lower and upper SOC-split bands are opposite of each other. 
The spin Rashba constant $\alpha_R$, quantifying the spin texture at $\Gamma$, for different materials are listed in Tab. \ref{tab:alpha}, and is compared to that of the literature. It suggests that the SRE is stronger in Te based Janus TMDs and  the largest $\alpha_R$ being for WSeTe. 
This further signifies the fact that while the orbital and spin magnetic moments are generated by the d-electrons of the transition metal atoms, the SOC effect is formed by the collective contribution from the transition metal atoms and the chalcogens. This is because the eigenstates that give rise to these magnetic moments are formed by a strong d-p hybridization.
\begin{table}[!htb]
\setlength{\tabcolsep}{6pt}
\renewcommand{\arraystretch}{1.1}

\begin{tabular}{l c|l c|l c}
\hline\hline
Material & $\alpha_R$ & Material & $\alpha_R$ & Material & $\alpha_R$ \\
\hline
MoSSe  & 75  & WSSe  & 157 & NbSSe  & 73  \\
MoSTe  & 152 & WSTe  & 320 & NbSTe  & 96  \\
MoSeTe & 482   & WSeTe & 526   & NbSeTe & 70   \\
\hline\hline
\end{tabular}

\caption{Spin Rashba constants ($\alpha_R$) in units of meV.\AA, measured at the $\Gamma$ valley for several Janus TMDs. The results are consistent with the previously reported values \cite{rashba-const}.  }

    \label{tab:alpha}

\end{table}
\begin{figure}[!htb]
\centering
\includegraphics[width=0.72\linewidth]{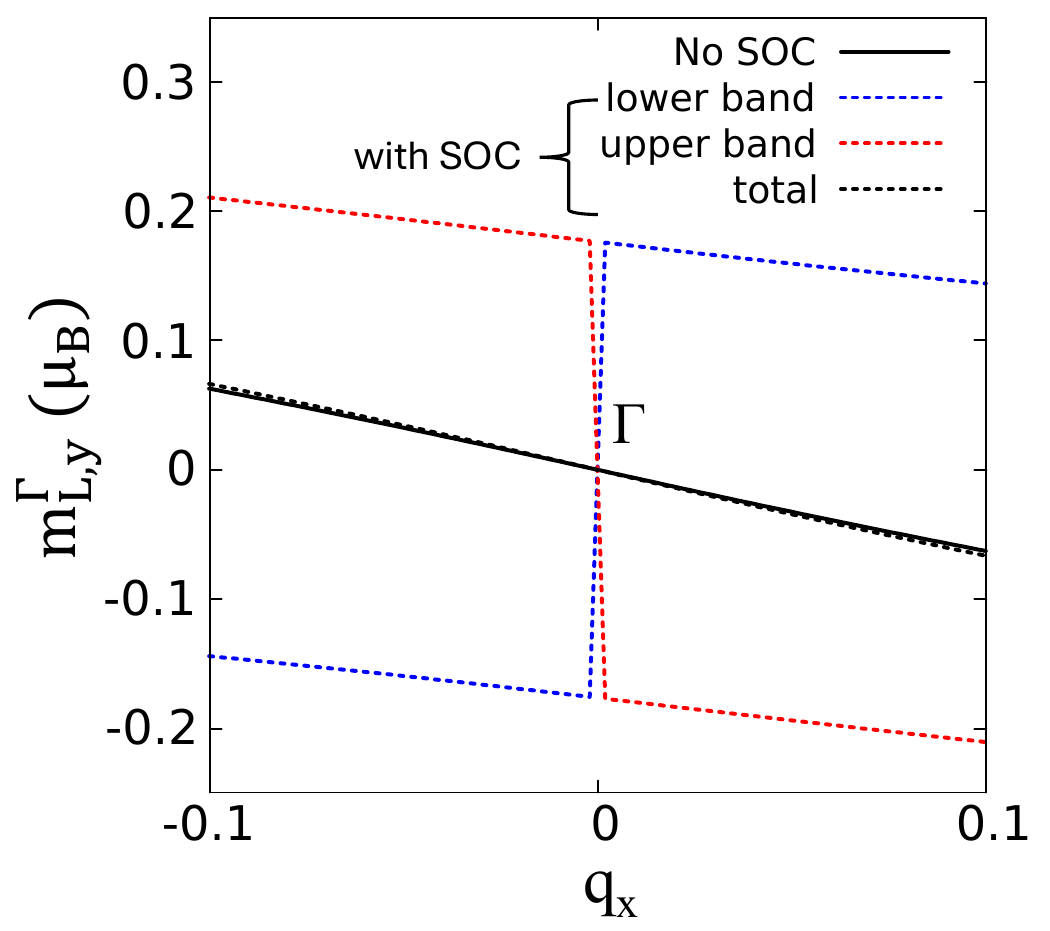}
\caption{Orbital moment $m_{L,y}^{\Gamma}$ along $q_x$ near the $\Gamma$ point. The black solid line presents the result for the case of no SOC, where the $m_L$ is formed by both the degenerate bands. In the presence of SOC, the contribution from upper and lower bands to the total $m_{L,y}$(black-dotted line) are shown with blue and red dotted lines. It shows that the total orbital moment remains unchanged even with the inclusion of SOC, although the upper and the lower bands show opposite values.} 
    \label{ly_qx_soc}
\end{figure}

To examine the cause of chirality reversal of the orbital texture, we turned our focus towards the independent role of the metal (M) and the chalcogen (X/X') atoms. For this purpose, we considered hypothetical cases in which the SOCs $\lambda_d$ of M and $\lambda_p$ of X/X$^{\prime}$ were turned off individually in the Wannier tight binding model. The results, shown in the supplemental material\cite{suppl}, infer that $\lambda_d$ has a negligible effect on the orbital texture. It tries to reverse the texture, but only in the immediate neighborhood of $\Gamma$. In contrast, due to the large $\lambda_p$ and strong p-d hybridization, the orbital moments align with the spin moments to create the chirality reversal. Hence, we see from Fig. \ref{fig:moments} that while the lower spin-split bands turn opposite with respect to the non-SOC orbital texture, the upper spin-split band maintains the same chirality as in the non-SOC case.
The opposite orbital textures of the upper and lower bands point towards a compensation of the total orbital moments which is quantified in Fig. \ref{ly_qx_soc}, where we plot the $m_{L,y}^{\Gamma}$ as a function of $q_x$. It shows that without the SOC, the degenerate bands have the same orbital moments(black solid line). However, as the SOC is introduced, the moments for the upper band (red dotted) changes from  positive to negative as q$_x$ changes from negative to positive, and the lower band (blue dotted) goes in an opposite manner. Despite this, we find the total orbital moment due to both the bands (black lines) to be the same with/without the SOC. Due to the chiral nature of the orbital texture, the same holds true in all other directions. Hence, when both bands are occupied, although the system does not have a net spin texture, the orbital texture still survives.
\section{Edelstein effect}

 \begin{figure*}[!htb]
 \includegraphics[width=\textwidth]{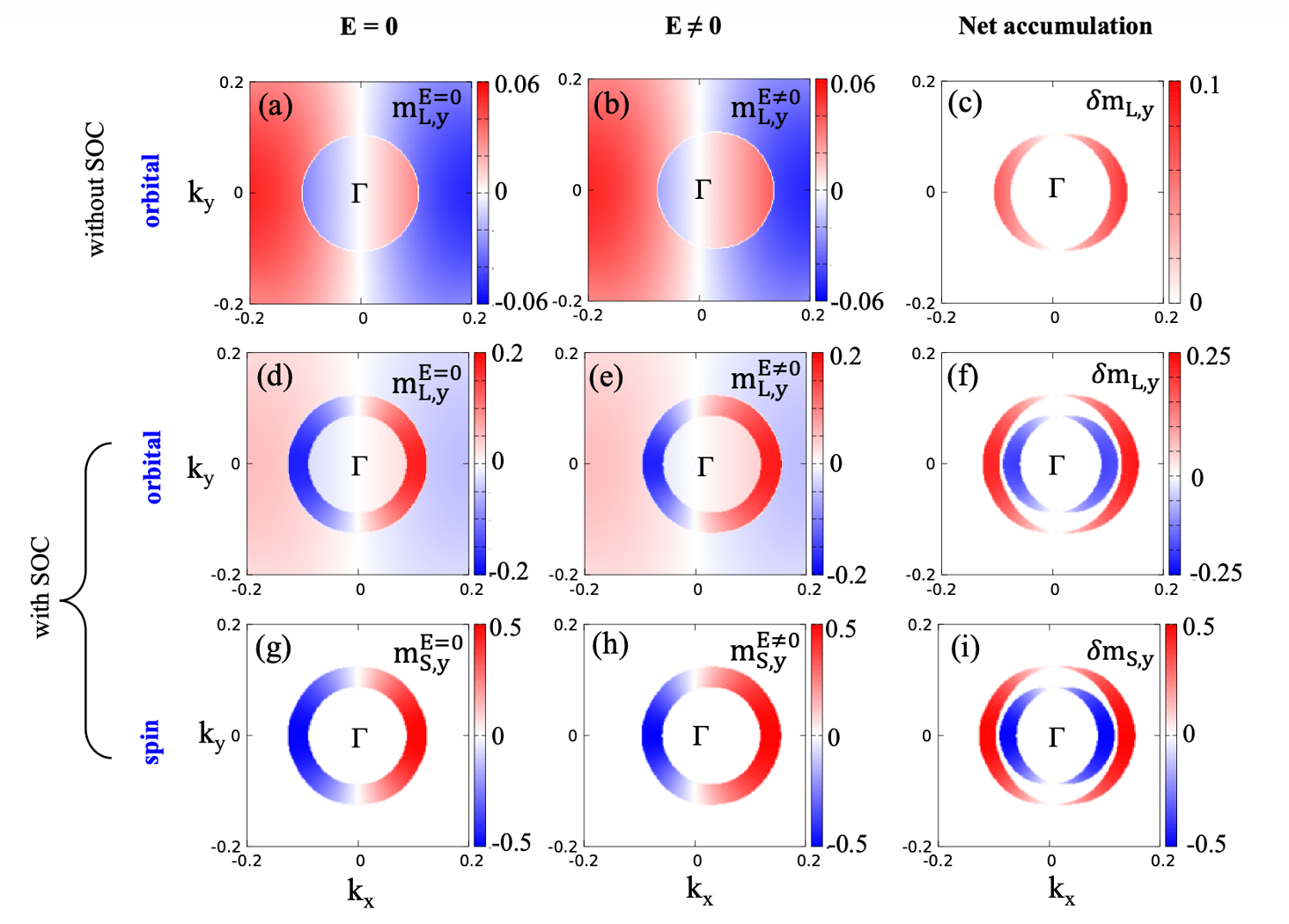}
   \caption{ Orbital magnetic moment for all occupied bands (without SOC) around $\Gamma$ point($m_{L,y}^\Gamma$), (a) without and (b) with the external electric field $E_x$. (c) Accumulation of orbital moment in the momentum space due to the fermi surface shift induced by the electric field without the SOC. The similar, in presence of SOC, is shown in (d) - (f) for the orbital moments, and (g) - (i) for the spin moments. Here, we  have considered a hole-doped case ($\varepsilon_F=-0.74$ eV$=\varepsilon_{VBM}-0.04$ eV).
   }
   \label{fig:edel-new}
\end{figure*}

The Rashba-Edelsteien effect, henceforth referred as the Edelstein effect, is the accumulation of magnetic moments across a sample in the presence of an in plane applied electric field ($\vec{E}$)\cite{edel-1}. Although a finite spin/orbital texture is observed due to the Rashba effect, the chiral nature of the distribution in the neighborhood of the valleys ensures that the net magnetization cancels out upon integration. However, with the application of $\vec{E}$, the Fermi surface shifts, which breaks the chirality of the magnetic moments and thereby induces a transverse magnetization in the sample, as schematically shown in Fig.\ref{fig:schematic}. We note that except for the Nb based compounds, which are metallic, the other Janus TMDs considered in this paper have a finite band gap. For such systems, it is necessary to have hole doping, which can be achieved either chemically or through bias, to create a Fermi surface, so that the Edelstein effect can be observed.

Considering $\vec{M}_{\zeta}$($\zeta=L$ or $S$) as the orbital/spin magnetization developed across the sample due to $\vec{E}$, and $\tilde{K}$ the orbital/spin Edelstein susceptibility tensor, the magneto-electric response can be expressed as:

\begin{eqnarray}
    \vec{M}_{\zeta} =\tilde{K}^{\zeta}\cdot \vec{E}.
    \label{edel1}
\end{eqnarray}
The net magnetization accumulated is $\vec{M}_{tot}  = \vec{M}_L+\vec{M}_S$.
To quantify this effect, first we calculate the Fermi surface shift due to the electric field. If $f_0(\varepsilon)$ is the Fermi surface without the electric field, and $\tau$ is the relaxation-time to account for the non equilibrium electron distribution, then the electric field induces a momentum shift of $\delta \vec{k} =-e\vec{E}\tau/\hbar$. So, the shifted Fermi surface can be expressed as

\begin{equation}
f(\varepsilon) = f_0(\varepsilon)+(\partial f/\partial k)\ \delta\vec{k} =  f_0-\frac{eE\tau}{\hbar}\frac{\partial f}{\partial \varepsilon}v(\vec{k}).
\end{equation}
Quantitatively, the spin/orbital magnetization can be obtained as follows. Taking the Fermi shift into account, the net change in magnetization under an applied electric field can be described as
\begin{equation}
    \begin{aligned}
        \vec{M}_{\zeta} & =  \frac{1}{(2\pi)^2}\int_{BZ}d^2k\ \vec{m}_\zeta(\vec{k})\times [f(\varepsilon_{k+\delta k}) - f(\varepsilon_k)], \\
    & = \frac{-eE\tau}{(2\pi)^2\hbar}\int_{BZ}d^2k \ \vec{m}_\zeta(\vec{k})\frac{\partial f}{\partial\varepsilon}v(\vec{k}).
    \end{aligned}
    \label{edel2}
\end{equation}
For simplicity, we have considered T=0. Therefore, Eq. \ref{edel2} becomes,

\begin{equation}
  \vec{M}_{\zeta} = \frac{-eE\tau}{(2\pi)^2\hbar}\int_{BZ}d^2k \ \vec{m}_\zeta(\vec{k})\delta(\varepsilon-\varepsilon_F)v(\vec{k}).
\end{equation}

The above equation states that only the magnetic moments at the Fermi surface   contribute to the Edelstein effect.
A visualization of the Edelstein effect from our numerical results for both OEE and SEE in MoSSe is shown in Fig. \ref{fig:edel-new}. The results are obtained for a hole doped case, with $\varepsilon_F = -0.74\ eV$($=\varepsilon_{VBM}-0.04$ eV). As can be observed in Fig.\ref{fig:edel-new}(a), (d), and (g), when the electric field is not applied, the moments $m^{E=0}_{L,y}(\vec{k})$ have opposite magnitudes along $\pm k_x$, and the symmetry of the distribution ensures that the net magnetization vanishes.  However, as the electric field $E_x$ is turned on(see Fig.\ref{fig:edel-new}(b), (e), and (h)), we notice a k-shift in the distribution, which makes $m^{E\neq0}_{L/S,y}(\vec{k})$ unequal for $\pm k_x$. Therefore, the accumulated moment distribution we get is $\delta m_{L/S,y} = m^{E\neq0}_{L/S,y}(\vec{k})-m^{E=0}_{L/S,y}(\vec{k})$, as shown in Fig. \ref{fig:edel-new}(c), (f), and (i). We notice that, when the SOC is absent, a positive ring is formed by $\delta m_{L}$, which leads to a net orbital magnetization, when integrated over the BZ.
\begin{figure}[!htb]
    \centering
    \includegraphics[width=0.5\linewidth]{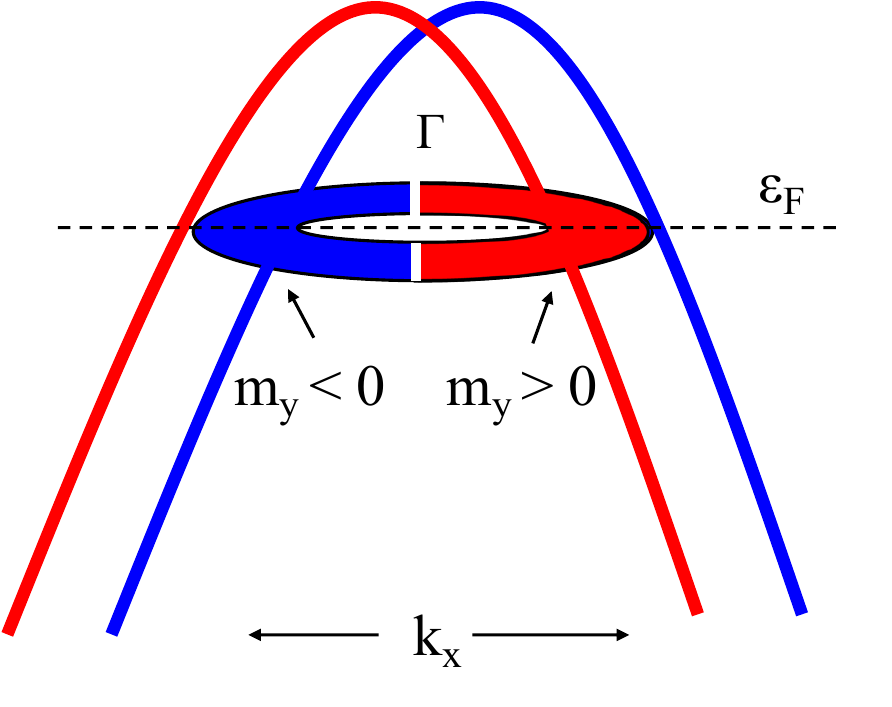}
    \caption{A schematic diagram showing the magnetic moments from all occupied bands for a hole doped system. The red and blue color indicates positive and negative moments respectively. The moments are only non-zero in the annular region, where only one of the bands is occupied. }
    \label{fig:mom_schematic}
\end{figure}

As we turn on the SOC, interesting phenomena occurs. The Rashba splitting in combination with the hole doping creates a ring shaped Fermi surface. The moments only survive in the region, where only one of the bands is occupied, as shown in the schematic Fig. \ref{fig:mom_schematic}. This explains the distribution of moments for $E=0$ in Fig. \ref{fig:edel-new}(d) and (g). 
Comparing the accumulated moments with the no-SOC case, we find that the positive $\delta m_{L}$ becomes stronger. However, due to the chirality reversal in the orbital texture, as discussed in the previous section, a relatively smaller negative $\delta_{m,L}$ ring inner to the positive $\delta_{m,L}$ ring emerges, as shown in Fig. \ref{fig:edel-new}(f). Additionally, due to the SOC, the SRE gives rise to spin Edelstein effect, and we observe similar positive and negative $\delta m_{S}$ rings, as shown in Fig.\ref{fig:edel-new}(e). We emphasize that these distribution can change as a function of $\varepsilon_F$. As more hole doping is introduced, the diameter of the rings formed by $\delta_{m,L}$ increases, and hence the thickness of the red and blue regions becomes relatively small, turning it into two thin circles. However, for the orbital moments, the region in between the thin circles can be non-zero for different occupations, due to the contribution from the occupied bands. The accumulation of the moments for a different hole doping is shown in the supplemental material\cite{suppl}.

\begin{figure}[!htb]
    \centering
    \includegraphics[width=0.72\linewidth]{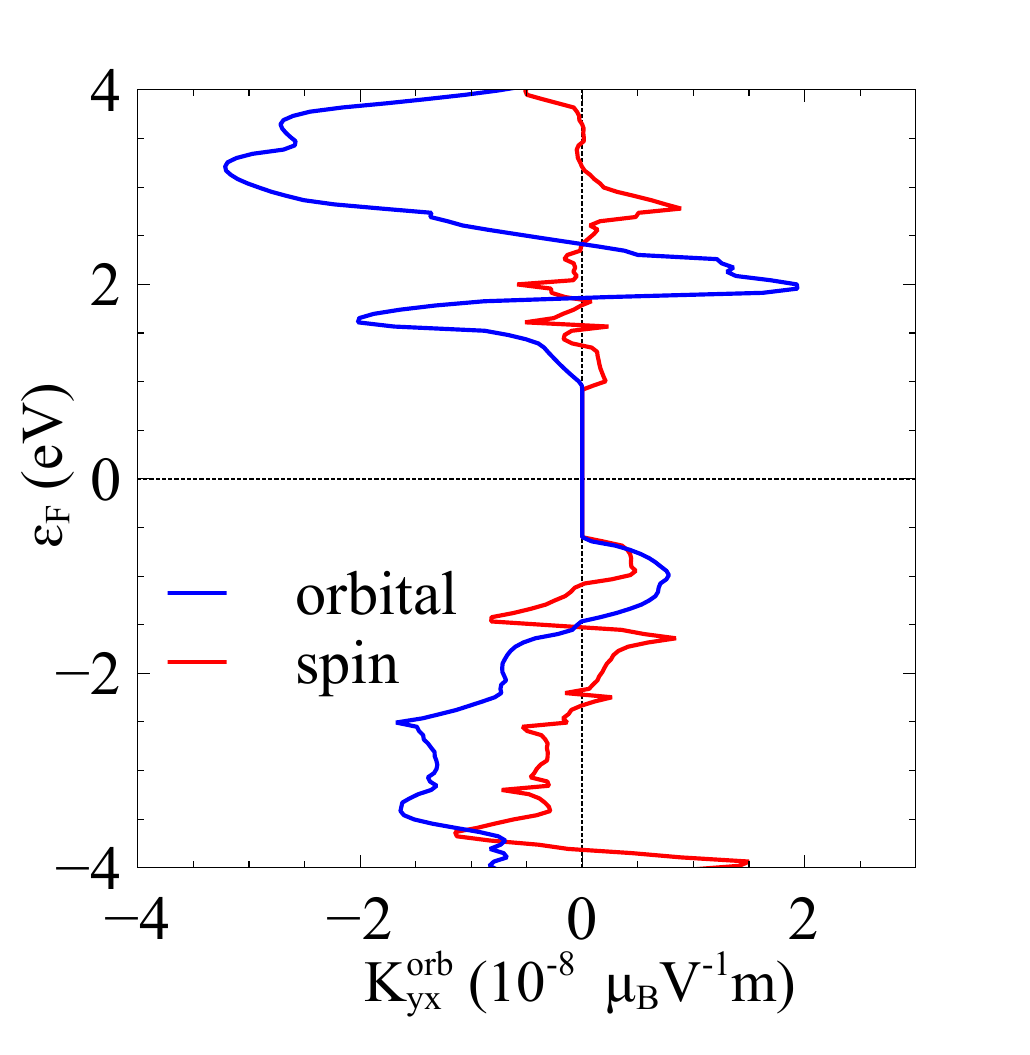}
    \caption{Orbital and spin Edelstein susceptibilities as a function of the occupation for MoSSe. The unit for $K_{yx}$ is $10^{-8}\mu_BV^{-1}m$. The OEE is stronger than the SEE throughout the range.}
    \label{fig:edel-mosse}
\end{figure}

Now to obtain the expression for the orbital/spin Edelstein susceptibility ($K_{ij}^\zeta$), we compare Eq. \ref{edel2} with Eq. \ref{edel1}, which yields

\begin{eqnarray}
    {K}_{ij}^{\zeta} = -\frac{e\tau}{(2\pi)^2\hbar} \int_{\text{BZ}} d^2k \, v_j(\vec{k}) \, \vec{m}_{\zeta}(\vec{k})\delta(\varepsilon-\varepsilon_F).
    \label{eq:chi}
\end{eqnarray}
Here, $i,j=x,y,z$, and the matrix element $K_{ij}^{\zeta}$ refers to the component of the orbital/spin susceptibility tensor, which measures the magnetization developed along $i$ for the applied electric field along $j$. Since Janus TMDs belong to the $C_{3v}$ point group, the susceptibility tensor has only two non-zero components, as predicted by symmetry\cite{edel-symm}, which are related to each other as $K_{xy}^{\zeta} = -{K_{yx}^{\zeta}}$.
We conveniently consider the electric field along $x$, and compute $K_{yx}^{\zeta}$ as a function of the Fermi energy $\varepsilon_F$. The integration is performed over the entire BZ in a 2D k-mesh grid of $300\times300$. The results for MoSSe are shown in Fig. \ref{fig:edel-mosse}. Since MoSSe is a semiconductor, both OEE and SEE are absent at $\varepsilon_F=0$. However, with a small hole doping, $K_{yx}$ takes a significantly high value, as shown in Fig.\ref{fig:edel-mosse}. Here, the interesting thing to note is that the OEE is much stronger SEE across the range of $\varepsilon_F$. We find this to be true for the rest of the materials of this class as well\cite{suppl}.  This suggests that a large accumulation of orbital moments can be observed in experiments with the application of an electric field, and the strength can be tuned by changing the occupation through the gate voltage.  Concerning  NbSSe and NbSTe, there is one less valence electron in the system, which makes them metallic, and therefore the finite Edelstein effects are observed in the pristine systems, which makes them ideal candidates to observe this effect experimentally.
\begin{figure}[!htb]
    \centering \includegraphics[width=0.45\textwidth]{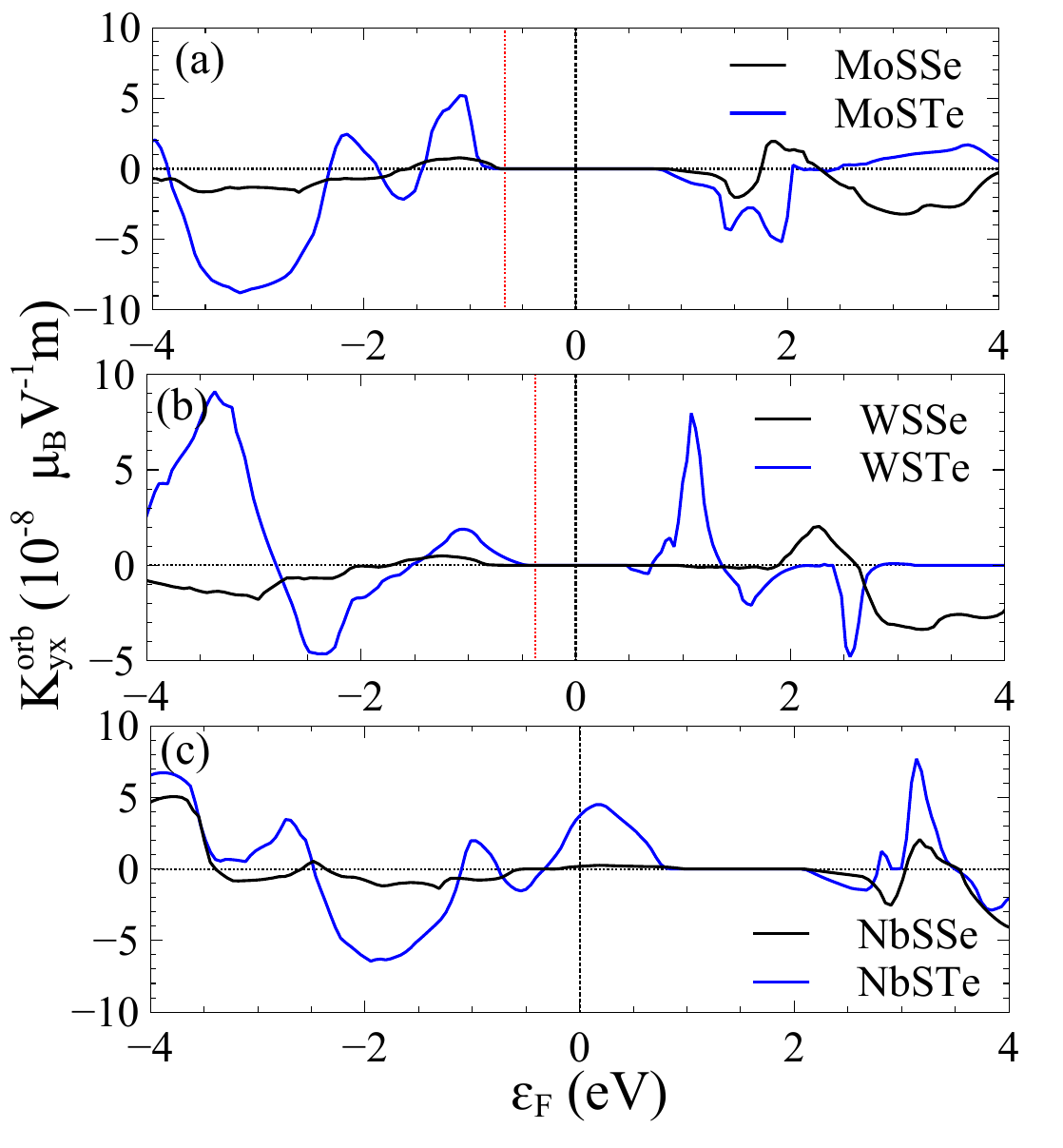}
    \caption{Orbital Edelstein susceptibility ($K_{yx}^{orb}$) as a function of $\varepsilon_F$ for MSSe and MSTe with latter showing large OEE. We attribute it to the strong $E_{int}$, when S and Te form the chalcogen layers. Here, the red dotted line shows the valence band maximum.
    }
    \label{fig:edel_compare}
\end{figure}

\begin{figure}[!htb]
    \centering
\includegraphics[width=0.87\linewidth]{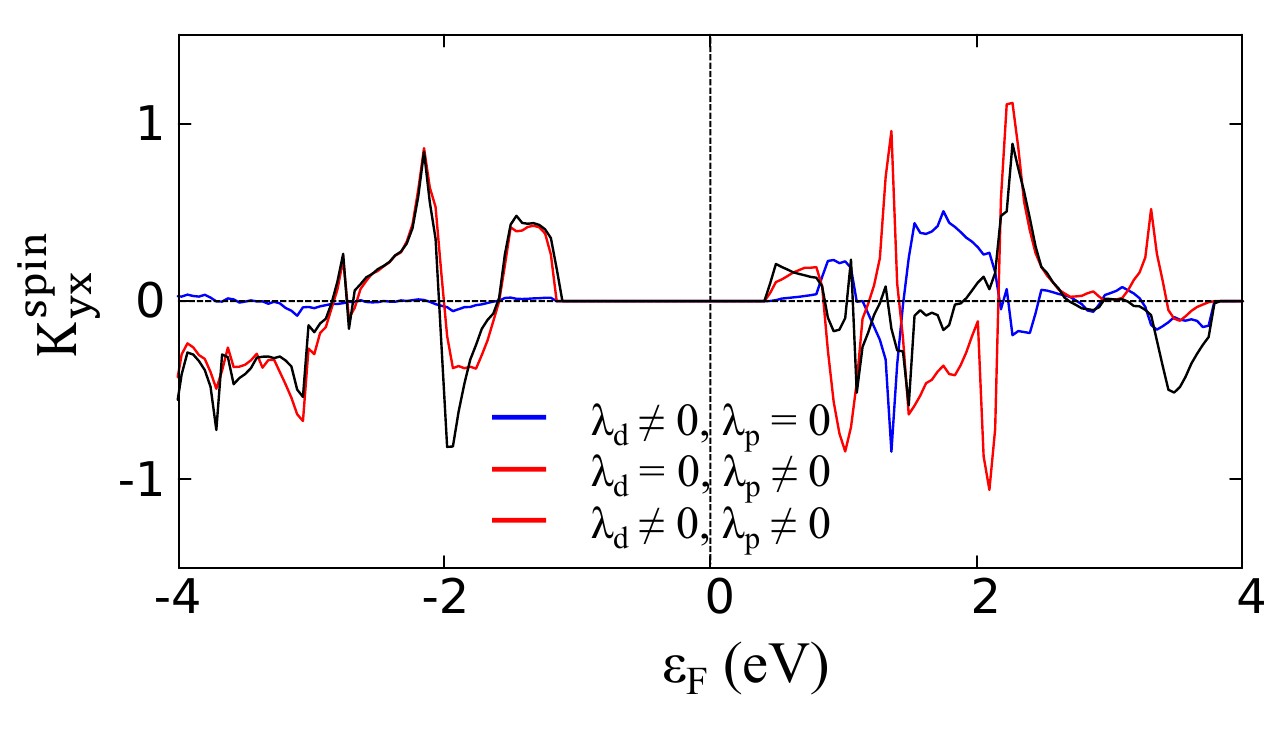}
    \caption{ Effect of SOC  ($\lambda_p$ and $\lambda_d$) on the Spin Edelstein susceptibility ($K_{yx}^{spin}$) throughout $\mu$ emphasizing a stronger effect of the $\lambda_p$ on the SEE.  }
    \label{fig:edel-soc}
\end{figure}

As discussed earlier, the Edelstein effect originates fundamentally from the Rashba effect, and hence the strength of mirror symmetry breaking strongly influences the magnitude of $K_{yx}^{\zeta}$. To further substantiate it, in Fig.\ref{fig:edel_compare}, we have calculated $K^{orb}$ for the pairs $MSSe$ and $MSTe$, since it is reported that $E_{int}$ is stronger for MSTe\cite{rashba-const}. From the figure, we see that the OEE is an order of magnitude stronger in MSTe as compared to MSSe. A similar trend is also observed for the SEE, which is shown in the Supplemental Material\cite{suppl}. In this case, along with the $E_{int}$, the strong SOC also contributes to the enhancement of the Edelstein effect. In fact, to examine the role of individual SOC on the SEE, we carried out a model study for different scenarios, where the SOC of both M, and $X/X^\prime$ are switched on, or only one of them is switched on, and the results are shown in Fig. \ref{fig:edel-soc} for MoSSe. From the figure, we gather that $\lambda_p$ of $X/X^\prime$  contributes the maximum to the SEE despite the spin and orbital textures are primarily formed by the d-orbitals. This signifies the role of strong p-d hybridization in shaping the eigenvectors.

\section{Summary}
In summary, we studied the orbital and spin textures in the monolayer Janus TMDs and examined the Eddelstein effect driven accumulation of spin and orbital magnetic moments in the presence of an applied in-plane electric field. We show that the momentum space spin and orbital texture in this class of materials arise due to the intrinsic electric field formed by the broken mirror symmetry. The intrinsic field, in addition to splitting the bands due to Rashba SOC at the vallyes $\Gamma$, K and K$^{\prime}$, alters the orbital composition of the band edges by inducing orbitals $\ket{xz}$ and $\ket{yz}$ in them. 
In a non-Janus TMD (MX$_2$) the valence and conduction band edges are formed by $\ket{x^2-y^2}$, $\ket{xy}$ and $\ket{z^2}$ orbitals. Finite orbital magnetic moment is obtained in the Janus system due to the addition of new orbital characters in the band edges which results in the formation of orbital texture. The Rashba SOC aligns the spin and orbital moments to induce a chirality reversal of the non-SOC orbital texture. As a consequence, we observe both orbital Edelstein (OEE) and spin Edelstein effect (SEE) due to the accumulation of the orbital and spin moments in the presence of an applied in-plane electric field. We further explored the magnetoelectric response by computing the Edelstein coefficient for both the orbital and spin components. The OEE is found to be linearly proportional to the Rashba SOC, and hence the Te based Janus materials exhibit maximum Edelstein susceptibility compared to the rest. Finally, note that while the OEE and SEE are naturally obtained for metallic Janus TMDs like NbSTe, carrier doping (holes or electrons) is required to realize the OEE/SEE in semiconducting Janus TMDs like MoSTe, WSSe, etc. The doping tunable OEE and SEE in Janus TMDs provide a promising platform for designing spin-orbitronic devices.

\section*{ACKNOWLEDGEMENT}
PS thanks IIT Madras for the financial support through institute of eminence (IoE) scheme. SS Thanks DST, India for the VAJRA fellowship.

\end{document}